\documentclass[twocolumn,aps,prl,amsmath,amssymb,showpacs,nofootinbib]{revtex4}
\usepackage{graphicx}
\usepackage{epsfig}

\def\beq{\begin{equation}}
\def\eeq{\end{equation}}

\newcommand{\kv}{\mathbf{k}}

\newcommand{\be}{\begin{eqnarray}}
\newcommand{\ee}{\end{eqnarray}}

\newcommand{\een}{\nonumber\end{eqnarray}}

\begin{document}
\thispagestyle{empty}
\vspace*{0.5 cm}
%\vspace*{-1.2in}
%\vspace*{0.7 in}
%\vspace*{-1.2in}
%\vspace*{0.7 in}
%\begin{center}
%title
\title{Nuclear matter from effective quark-quark interaction.}
%\vspace*{1cm} 
\author{\bf M. Baldo and K. Fukukawa}  
%\vspace*{.3cm}
\affiliation{INFN, Sezione di Catania, via S. Sofia 64, I-95123, Catania, Italy}
%\vspace*{.3cm}
%

%\vskip -0.5 cm
%\end{center}
\begin{abstract}
\noindent We study neutron matter and symmetric nuclear matter with the quark-meson model for the two-nucleon interaction. The Bethe-Bruckner-Goldstone
many-body theory is used to describe the correlations up to the three hole-line approximation with no extra parameters.
At variance with other non-relativistic realistic interactions, the three hole-line contribution turns out to be non-negligible and to have a substantial saturation effect. The saturation point of nuclear matter, the compressibility, the symmetry energy and its slope are within the phenomenological constraints. Since the interaction also reproduces fairly well the properties of the three nucleon system, these results indicate that the explicit introduction of the quark degrees of freedom within the considered constituent quark model is expected to reduce the role of three-body forces.   
\end{abstract}               
%\vskip -0.4 cm
%\noindent
%\begin{flushleft}
%\end{flushleft}
%24.10.Cn ,  % Many body theory
%26.60.+c ,  % Nuclear matter aspects of neutron stars
%03.75.Ss    % Degenerate Fermi gases
%\vfill\eject

\maketitle
%%%%%%%%%%%%%%%%%%%%%%%%%%%%%%%%%%%%%%
%%%%%%%%%%%%%%%%%%%%%%%%%%%%%%%%%%%%%%
%%%%%%%%%%%%%%%%%%%%%%%%%%%%%%%%%%%%%%
%%%%%%%%%%%%%%%%%%%%%%%%%%%%%%%%%%%%%%
%%%%%%%%%%%%%%%%%%%%%%%%%%%%%%%%%%%%%%
%%%%%%%%%%%%%%%%%%%%%%%%%%%%%%%%%%%%%%

\par\noindent
{\it Introduction.}
Understanding the properties of the nuclear medium on the basis of the bare interaction among nucleons is one of the fundamental issue of Nuclear Physics. Several methods have been used to model the nucleon interaction and to develop accurate many-body theory to describe the correlations in nuclear systems.
It has been established \cite{Day,Song,Bal1} that realistic nucleon-nucleon (NN) potentials based on the meson-exchange interaction model fail to reproduce the correct saturation point and require the introduction of three-body forces (TBF). The latter can be phenomenological \cite{Urb,Tar} or more fundamental \cite{Umb,Hans}. In any case, within this framework, the effect of TBF is moderate, but it is essential to shift the saturation point inside the phenomenological boundaries. The main problem of this approach is that it appears difficult to device a TBF that describes satisfactorily well few-body systems and at the same time nuclear matter near saturation \cite{Wir}. The same conclusion has been reached within the variational many-body method \cite{APR}. In the framework of this type of nuclear forces, non-local two-body interactions \cite{Dolles} have been constructed that reproduce closely the binding energy of three and four nuclear systems. However, they fail to reproduce the correct saturation point \cite{Chiara1}. The Dirac-Brueckner-Hartree-Fock (DBHF) method introduces relativistic effects in the many-body theory. With only two-body forces and two-body correlations, the saturation point is fairly well reproduced \cite{Fuchs}, but the problem of the few-body systems remains unsolved. It can also be shown \cite{Brown} that the relativistic effects introduced by the DBHF scheme are equivalent to a particular TBF at the non-relativistic level. 
More recently the chiral effective forces have been developed \cite{Weinberg,Ent}. These forces were devised to connect the underlying Quantum Chromodynamics (QCD) theory of strong interaction among quarks to the low energy interaction among nucleons. However no explicit quark degrees of freedom are introduced, but it is based on the expansion of the interaction in the chiral symmetry breaking parameter, i.e. the $\rho$-meson mass $m_\rho$. As such, it is an expansion, in $k/m_\rho$, where $k$ is the typical nucleon momentum. 
It has the fundamental property to classify the forces according to the expected relevance, following a "power counting" rule. Three-body (or higher) forces arise naturally in this expansion. Even if the assignment of the order to the different
 interaction processes looks tricky \cite{Valder}, this approach has been developed both at the fundamental level \cite{Leut,Ulf,Epel} and in a wealth of applications to nuclei \cite{Ca} and nuclear matter \cite{Hebel,Diri,Hebel2,Rios}. The parameters of the forces are fixed by fitting the NN phase shifts and eventually the properties of the three-body system.
 However the procedure looks not unique. In ref. \cite{Mac} it has been shown that it is possible to construct a realistic chiral two-body force that reproduces the spectroscopic data on light nuclei without invoking TBF. However symmetric nuclear matter was not considered.  
  In ref. \cite{MacCor} it has been shown that a version of chiral force was able to reproduce, by a suitable choice of the momentum cut-off parameter, the few-body binding energies and at the same time a fair saturation point. However, this remarkable result needs confirmation, since two and three-body forces were actually taken at different orders and some correlation diagrams were neglected. 
 In any case, for all chiral forces TBF are the dominant mechanism for saturation. Indeed, with only two-body forces no saturation is apparent in nuclear matter.  
% A different approach has been followed by Fujiwara and collaborators \cite{Fuji}. In this case the quark %degrees of freedom are explicitly introduced and the nucleon-nucleon force is derived from 
%the quark-quark interactions. The latter is an effective meson-exchange potential between quarks. 
%The Resonant Group Method is used to deal with the effect of the quark wave function inside each %nucleons. The resulting interaction is highly non-local and contains a natural cut-off in momentum. More %details on this Quark Model (QM) force can be found in ref. \cite{Kenji}.
%\par
NN potentials based on the constituent quark model have been developed for some decades,
since the resonating-group-method (RGM) equations was firstly solved by Oka and Yazaki \cite{OY8081}.
In this model the quark degrees of freedom are explicitly introduced
and the NN potential is derived from the quark-quark (qq) interactions.
The resulting interaction is highly non-local due to the RGM formalism and contains a natural cut-off in momentum.
Realistic quark-model (QM) interactions were proposed by Fujiwara and collaborators \cite{Fuji},
in which the qq interaction consists of a color-analog of the Fermi-Breit interaction
and an effective meson-exchange potential.
The most recent model fss2  \cite{fss2NNYN} reproduces the experimental data on the few-body systems
(triton, hypertriton \cite{hypt08} and $\alpha$ particle \cite{alpha} and nucleon-deuteron scattering
\cite{ndscatreview}) fairly well without introducing TBF. 
 In this letter we present results for nuclear matter obtained with the QM force within the Bethe-Bruekner-Goldstone (BBG) many-body expansion up to three hole-line level of approximation.
 A pedagogical introduction to this many-body method can be found in ref. \cite{book}.
The energy dependence inherent to the RGM formalism is eliminated by the off-shell transformation
utilizing the norm kernel as in Refs. \cite{Suzuki, ndscat1} and
the Gaussian representation of fss2  \cite{KFMPLA} is used.              
The set of Goldstone diagrams that are used in the calculations is reported in Fig. 1. 
\begin{figure}[t]
\vspace {-2 cm}
\begin{center}
\includegraphics[width=1.008\linewidth,height=1.188\linewidth]{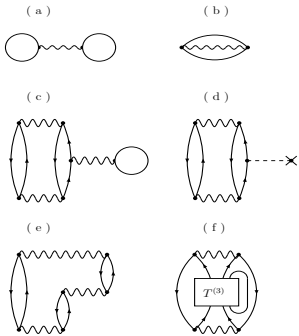}
%\vspace{-6 cm}
\begin{minipage}[c]{0.9\linewidth}
\vspace{-6. cm}
\caption{Different Goldstone diagrams contributing to the Nuclear Matter EOS. The wavy line indicates Brueckner G-matrix. The box labelled $T^{(3)}$ is the in-medium three-body scattering matrix. Diagram (c) is the first term of the set of diagrams obtained once the expansion of $T^{(3)}$ is inserted in (f) and it has been singled out for numerical convenience.} \label{fig1}
\end{minipage}
\vspace{-2.3 cm}
\end{center}
\end{figure}
%%%%%%%%%%%%%%%%%%%%%%%%%%%%%%%%%%%%%%%%%%%%%%%%%%%%%%%%%%%%%%%
\begin{table}
\caption{Three hole-line contributions to the symmetric matter EOS for different
Fermi momenta $k_F$ in fm$^{-1}$. $E_3$ is the
total three hole-line contribution, $B$ is the ``bubble diagram" of Fig. 1(c),
BU is the $U$-insertion diagram of Fig. 1(d), R is the ``ring diagram" of Fig.
1(e) and H indicates the ``higher order" diagrams, as defined in the text.
Energies are in MeV. }
\begin{ruledtabular}
\begin{tabular}{cccccccc}

 $ k_{F}$ & $T+E_2$ & $B$ & $BU$ &  $R$  & $H$ & $E_3$ & $EOS$  \\
\hline
  1.1 & -17.090 &  -7.020  & 10.655 & -0.750 & 0.177 & 3.072 & -14.028  \\
  1.2 & -19.680  & -6.351 & 11.407 & -1.270 & 0.157 & 3.943 & -15.737  \\
  1.3 & -22.154&  -4.669  & 11.647 & -1.761 & 0.144 & 5.361 & -16.793  \\
  1.4 & -24.393 &  -2.689  & 12.340 & -2.030 & -0.079 & 7.542 & -16.851  \\
  1.5 & -26.183 &  0.223  & 12.781 & -2.122 & 0.050 & 11.022 & -15.161  \\
  1.6 & -27.498 &  4.162  & 13.759 & -2.280 & 0.029 & 15.670 & -11.828  \\
\end{tabular}
\end{ruledtabular}
\vskip -0.2 cm
\end{table}
The BBG expansion classifies the diagrams according to the number of hole-lines that they contain. Diagrams (a) and (b)
(direct and exchange) include two hole-lines and they correspond to the well-known Brueckner-Hartree-Fock (BHF) approximation. The wavy line indicates the Brueckner G-matrix \cite{book}.
In the BBG expansion an auxiliary single-particle potential $U(\kv)$ is introduced, and calculated self-consistently according to the Brueckner prescription. However the auxiliary potential is not unique. Two somehow opposite choices are possible. In the so called ''standard'' or gap choice (GC), the potential is assumed to be zero above the Fermi momentum, while in the ''continuous'' choice (CC) the potential is calculated self-consistently also above the Fermi momentum. In principle the final result should be independent of $U$, which is introduced in order to rearrange the perturbation series for a faster convergence.
The comparison of the result obtained with the gap and the continuous choice can be used to estimate the degree of convergence of the expansion \cite{Song,Bal1}. The rearrangement of the expansion is embodied in   
a series of "$U$-insertion" diagrams. The first one is diagram (d), while diagram (c) is a self-energy insertion. Notice that the various self-energy or $U$-insertion diagrams must follow from the BBG expansion, otherwise an arbitrary number of insertions would spoil the hole-line ordering of the diagrams. Diagram (e) is generally indicated as "ring diagram". It describes long range correlations in the matter. Diagram (f) describes the full scattering process of three particles that are virtually excited above the Fermi sphere and its evaluation requires the solution of the Bethe-Fadeev equations \cite{Bethe,Day,book}.  
The sum of the diagrams (c-f) gives the three hole-line contribution, which, according to the BBG hole-line expansion, is expected to be substantially smaller than the two hole-line (Brueckner) contribution.       
\par\noindent
{\it Results.}
The results for symmetric matter (SM) in the CC are reported in Table I for a set of values of Fermi momenta around saturation, with the breakdown of the contributions of each diagram. 
\par
%In this letter we focus on the EOS around saturation density, at Fermi momenta $ 1.1 \, \le \, k_F \le %1.6$ fm$^{-1} $. 
The last column reports the final EOS obtained summing up the two hole and three hole-line contributions. In the column before the last the total contribution of the three hole-line diagrams is reported. 
%%%%%%%%%%%%%%%%%%%%%%%%%%%%%%%%%%%%%%%%%%%%%%%%%%%%%%%%%%%%%%%
\begin{table}
\caption{The same as in Table I, but in the gap choice for the single particle potential. }
\begin{ruledtabular}
\begin{tabular}{cccccccc}

 $ k_{F}$ & $T+E_2$ & $B$  &  $R$  & $H$ & $E_3$ & $EOS$  \\
\hline
  1.1 & -11.605 & -0.556   & -1.003 &  0.063 & -1.496 & -13.101 &  \\
  1.2 & -13.525 & -0.029   & -1.119 &  0.040 & -1.108 & -14.633 &  \\
  1.3 & -15.439 &  0.846   & -1.251 &  0.033 & -0.372 & -15.721 &  \\
  1.4 & -16.959 &  2.213   & -1.301 &  0.021 &  0.933 & -16.026 &  \\
  1.5 & -18.212 &  4.234   & -1.296 &  0.012 &  2.575 & -15.272 &  \\
  1.6 & -18.974 &  7.233   & -1.328 &  0.006 &  5.911 & -13.063 &  \\
\end{tabular}
\end{ruledtabular}
\vskip -0.2 cm
\end{table}
Table II is similar but for the gap choice. Notice that in this case the $U$-insertion diagram (d) vanishes. The comparison between the two sets of results for the EOS of SM is summarized in Fig. 2 for a wider range of density. 
\begin{figure}[h]
\begin{center}
\includegraphics[width=0.72\linewidth,height=0.504\linewidth]{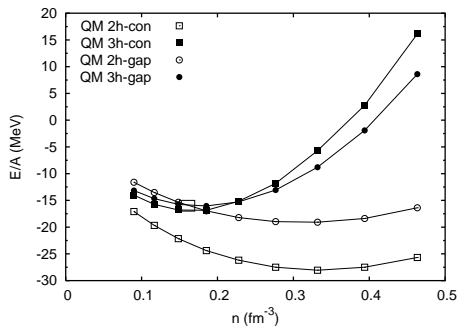}
\begin{minipage}[c]{0.9\linewidth}
\caption{EOS of symmetric Nuclear Matter at the two-hole level of approximation (open symbols) and at three hole-line level (full symbols) for the continuous (squares) and the gap (circles) choices, respectively.} \label{fig2}
\end{minipage}
\end{center}
\vskip -0.6 cm
\end{figure}
%%%%%%%%%%%%%%%%%%%%%%%%%%%%%%%%%%%%%%%%%%%%%%%%%%%%%%%%%%%%%%%
One can see that at the Brueckner (two hole-lines) level of approximation the continuous and gap choices differ by a few MeV, being the continuous one more attractive. However, as the three hole-line contribution is added, the two EOS are quite close. The discrepancy around saturation is not exceeding 1 MeV and it is vanishing small just at saturation. We consider this result as a strong indication of the convergence of the BBG expansion. However, at high density the discrepancy tends to increase, which indicates a lower degree of convergence, but no divergence of the expansion is really apparent. It has to be noticed that for other NN interactions, in particular the Argonne v$_{18}$ potential, and the corresponding simplified versions v$_{8}$, v$_{6}$ and v$_{4}$, the three hole-line contribution is much smaller in the CC than in the GC \cite{gang}. For the present QM potential it is the opposite, in the gap choice
the convergence of the energy looks faster.
% However, it has to be kept in mind that the continuous choice appears more physical, since the potential %has no discontinuity, and it can be more appropriate for the calculations of phenomenological quantities %like the optical potential. 
 One can notice that around saturation the ratio between the three hole-line correlation energy and the two hole-line one (BHF) is 0.15 for the CC and 0.02 for the GC. This is in line with the expectation of the hole-line expansion and supports the validity of the BBG expansion. Notice that the second columns of Tables I and II include the free kinetic energy $T$.   
If the EOS around saturation density are fitted with a form of the type $ E/A \, =\, a\rho \, +\, b\rho^\gamma  $ the saturation point turns out to be $e_0 = -16.9$ MeV and $\rho_0 = 0.166$ fm $^{-3}$ for the CC and $e_0 = -16.06$ MeV and $\rho_0 = 0.177$ fm $^{-3}$ for the GC. This establishes the range of the uncertainty on the predicted saturation point and of the whole EOS in the considered density range. From the same fits one can extract the compressibility at saturation, which turns out to be
$ K \,=\, 228$ MeV  for the CC and $ K \,=\, 192 $ MeV for the GC. These values can be considered  compatible with the range encompassed by the phenomenological constraints \cite{Dutra}, the GC value being at the lower edge. 
A similar analysis can be performed for pure Neutron Matter (PNM). The corresponding EOS for the CC and GC are reported in Fig. 3. 
\begin{figure}[h]
\begin{center}
\includegraphics[width=0.72\linewidth,height=0.504\linewidth]{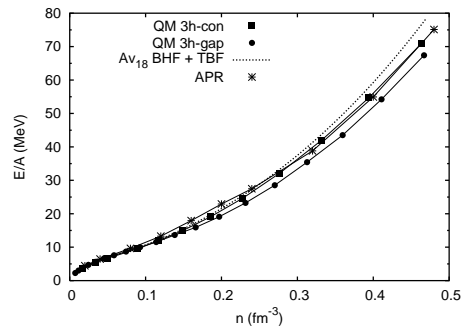}
\begin{minipage}[c]{0.9\linewidth}
\caption{The EOS of pure Neutron Matter at the three-hole line level of approximation with the QM potential in the continuous (full squares) and gap (full circles) choices. For comparison one BHF EOS from ref. \cite{Tar} (dashed line) and the one of ref. \cite{APR} (stars) are also reported. The latter two includes three-body forces} \label{fig3}
\end{minipage}
\end{center}
\end{figure}
\vskip 0.2 cm
%%%%%%%%%%%%%%%%%%%%%%%%%%%%%%%%%%%%%%%%%%%%%%%%%%%%%%%%%%%%%%%
\begin{figure}[h]
\begin{center}
\vskip -0.4 cm
\includegraphics[width=0.72\linewidth,height=0.504\linewidth]{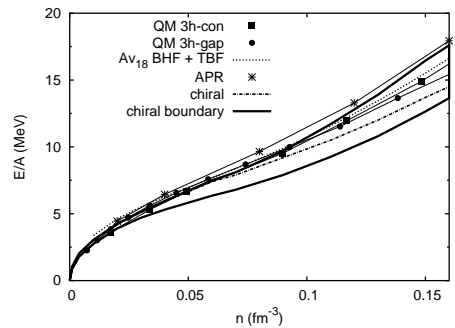}
\begin{minipage}[c]{0.9\linewidth}
\caption{Blow up of Fig. 3 in the low density region. For comparison the EOS from the chiral expansion 
of ref. \cite{MacCor2} (dashed-dotted line) has been reported. The area delimited by the thick full line indicates the region where the EOS predicted from the chiral approach of ref. \cite{Tews} should be enclosed.   } \label{fig4}
\end{minipage}
\end{center}
\vskip -0.6 cm
\end{figure}
\vskip -0.4 cm
%%%%%%%%%%%%%%%%%%%%%%%%%%%%%%%%%%%%%%%%%%%%%%%%%%%%%%%%%%%%%%%
%The three hole-line contributions turn out to be substantially smaller than in the case of SM. 
For comparison two EOS which include three-body forces are also reported, one from the BHF approach \cite{Tar} and one from the variational method \cite{APR}. A blow up of the low density region is reported in Fig. 4, where in addition the EOS obtained from the chiral force approach of ref. \cite{MacCor2} is reported. The region enclosed inside the thick full line indicates the allowed area where the EOS should pass through according to the chiral approach of ref. \cite{Tews}. The relatively close agreement among the different approaches shows that the separation between two-body and three-body forces, as well as the relevance of higher order correlations, are substantially model dependent.  
The symmetry energy $S(\rho)$ can be then extracted as difference between PNM and SM, which is valid for a quadratic dependence of the EOS on asymmetry. The fits give also the derivative of asymmetry at saturation, as embodied in the parameter $L = 3\rho (\partial S/\partial \rho)$. One finds $S_0 \,=\, 34$ MeV and $L = 54$ MeV for the CC, and $S_0 \,=\, 33.7$ MeV and $L = 53$ MeV or the GC, again compatible with phenomenology \cite{Dutra}.
\par\noindent
{\it Discussion.} The microscopic EOS obtained from the QM interaction compares well with phenomenological constraints, at variance with modern NN meson exchange interaction models.  
Since the interactions are phase equivalent, the reason of the discrepancy must be due to the different off-shell behaviour of the QM interaction. In particular, this can be related to the characteristic non-locality of the repulsive core \cite{ndscatreview} as produced by the quark exchange processes. Notice that the presence of non-locality not necessarily improves the saturation point \cite{Chiara1}. We also found that at the BHF level the contribution of the
$^3S_1 - ^3D_1$ channel is much larger than for the other interactions, e.g. the $Av_{18}$.
% despite the regular value of the the D-wave percentage ( 6.6 \% ) in deuteron.
This large contribution is responsible almost completely for the too attractive EOS in BHF, whose saturation point is well outside the Coester band. 
The larger  three hole-line contribution for QM than for $Av_{18}$ is qualitatively in line with the trend of the corresponding values of the wound parameters at saturation, 0.072 and 0.037, respectively. However a large contribution (0.019) to the wound parameter comes from the $^3S_1 - ^3D_1$ channel in $Av_{18}$, while this is quite small for QM (0.009), probably due to the softer short range repulsion. A comparison with $Av_{18}$
shows that the saturation mechanism with the QM is the steeper increase with density of the 3hole-line contribution. This is due to two reasons, i) The larger value of the $U$-insertion diagrams
and its increase with density, while for $Av_{18}$ it is decreasing, ii) the very small values of the higher order diagrams, which in $Av_{18}$ are relevant and negative.     
    
%We have presented the calculation of the EOS both for symmetric matter and pure neutron matter based on %the NN interaction by Fujiwara and collaborators. Since the interaction is derived from an effective %quark-quark interaction within the RGM, the EOS are directly related to the QCD inner structure of the %nucleons. The BBG many-body theory has been used up to the three hole-line level of approximation. The %expansion shows a good degree of convergence, and the saturation point, the compressibility, the symmetry %energy and its derivative compare well with the phenomenological data, with an error that is within the %phenomenological uncertainty. The same NN interaction is able to reproduce fairly well the $^3H$ binding %\cite{hypt08} and the scattering data on proton-deuteron system \cite{ndscatreview}. These results %suggest that, if the NN interaction is modelled from the quark-quark interaction, the need of three-body %forces is reduced to a minimum. 
The need of three-body forces with the QM interaction seems to be reduced to a minimum.
The consistency of this conclusion could be checked 
by deriving explicitly the three-body forces from the same quark model, which is left to a future long term project.  
\vskip 0.2 cm
\par\noindent Partial support from NewCompStar, COST Action MP1304 is gratefully acknowledged.    

\end{document}